\documentclass[aps,pre,amsmath,onecolumn,showpacs]{revtex4-2}
\usepackage{amsmath}
\usepackage{amsfonts}
\usepackage{amssymb}
\usepackage{graphicx}
\usepackage{hyperref}
\usepackage[dvipsnames,table]{xcolor}
\usepackage{mathtools}
\usepackage{verbatim}
\usepackage{epstopdf}
\usepackage{subfigure}
\usepackage{latexsym}
\usepackage{dcolumn}
\usepackage{epsf}
\usepackage{float}
\usepackage{multirow}
\usepackage[utf8]{inputenc}
\allowdisplaybreaks[1]
\usepackage{ulem}
\usepackage{amsmath}
\usepackage{amsfonts}
\usepackage{amssymb}
\usepackage{graphicx}
\usepackage{hyperref}
\usepackage{mathtools}
\usepackage{verbatim}
\usepackage{epstopdf}
\usepackage{subfigure}
\usepackage{latexsym}
\usepackage{dcolumn}
\usepackage{epsf}
\usepackage{float}
\usepackage{graphicx}
\usepackage{dcolumn}
\usepackage{color}
\usepackage{graphicx}
\usepackage{amssymb,amsmath}
\usepackage{sidecap}
\usepackage{bm}
\usepackage{color,soul}
\definecolor{chocolate(traditional)}{rgb}{0.48, 0.25, 0.0}
\definecolor{coquelicot}{rgb}{1.0, 0.22, 0.0}
\usepackage{color}
\begin{document}
	
		\newcommand{\I} 
	{
		\mathcal I
	}
\title{Enhancing cluster synchronization in phase-lagged multilayer networks}

\author{Abhijit Mondal$^{1,*}$}
\author{Pitambar Khanra$^{2,*,**}$}
\author{Subrata Ghosh $^{3,4}$}
\author{Prosenjit Kundu$^5$}

\author{Chittaranjan Hens$^3$}
\author{Pinaki Pal$^1$}
\affiliation{$^1$Department of Mathematics, National Institute of Technology, Durgapur~713209, India}
\affiliation{$^2$School of Computer Science, UPES, Dehradun, Uttarakhand~248007, India}
\affiliation{$^3$Center for Computational Natural Science and Bioinformatics, International Institute of Informational Technology, Gachibowli, Hyderabad 500032, India}
\affiliation{$^4$Division of Dynamics,  Lodz University of Technology, Stefanowskiego 1/15, 90-924  Lodz, Poland}
\affiliation{$^5$Dhirubhai Ambani Institute of Information and Communication Technology, Gandhinagar, Gujarat, 382007, India}
\affiliation{$^*$Contributed Equally}
\affiliation{$^{**}$Corresponding Author (pitambar.khanra@ddn.upes.ac.in)}

\begin{abstract}
Cluster synchronization in multilayer networks of phase oscillators with phase-lag poses significant challenges due to the destabilizing effects of delayed interactions. Leveraging the Sakaguchi-Kuramoto model, this study addresses these challenges by systematically exploring the role of natural frequency distributions in sustaining { cluster synchronization} under high phase-lag conditions. We focus on four distributions: uniform (uni-uni), partially degree-correlated (deg-uni, uni-deg), and fully degree-correlated (deg-deg), where oscillators’ intrinsic frequencies align with their network connectivity. Through numerical and analytical investigations, we demonstrate that the deg-deg distribution where both layers employ degree-matched frequencies remarkably enhances synchronization stability, outperforming other configurations. We analyze two distinct network architectures: one composed entirely of nontrivial clusters and another combining trivial and nontrivial clusters. Results reveal that structural heterogeneity encoded in the deg-deg coupling counteracts phase-lag-induced desynchronization, enabling robust { cluster synchronization} even at large phase-lag values. Stability is rigorously validated via transverse Lyapunov exponents (TLEs), which confirm that deg-deg networks exhibit broader synchronization regimes compared to uniform or partially correlated systems. These findings provide critical insights into the interplay between topological heterogeneity and dynamical resilience, offering a framework for designing robust multilayer systems from delay-tolerant power grids to adaptive biological networks, where synchronization under phase-lag is paramount.
\end{abstract}
\maketitle
\section{Introduction}
Complex systems, whether natural or man-made, often consist of interacting dynamical units and are prevalent in various contexts. Examples include the brain~\cite{adhikari2013localizing}, where synchronized neuronal oscillations underpin cognitive functions; power grids~\cite{buldyrev2010catastrophic}, whose stability relies on the coordinated dynamics of generators and loads; groups of fireflies~\cite{buck1988synchronous}, which flash in unison through mutual entrainment; and flocks of birds~\cite{attanasi2015emergence}, whose collective motion emerges from local interactions. To mathematically model the structure of such systems, researchers employ graphs or networks, where vertices represent dynamical units and edges denote interactions. While simple graphs suffice for some systems, real-world scenarios-such as power grids, transportation networks, neuronal circuits, social interactions, and epidemic spreading~\cite{granell2013dynamical,nicosia2017collective,danziger2019dynamic,soriano2019explosive,de2016physics,soriano2018spreading} often involve multilayer architectures, where distinct interaction types coexist across layers. Critically, the intrinsic dynamics of these units must be integrated with the network topology to fully unravel system behavior. Among the myriad dynamical phenomena, synchronization-the alignment of oscillators’ rhythms-stands out as a cornerstone of collective behavior.

The Kuramoto model~\cite{rodrigues2016kuramoto,skardal2018low} has been instrumental in studying synchronization, revealing how global synchronization emerges when oscillators align their phases through coupling. { Global synchronization} depends on network structure and the master stability function~\cite{pecora1998master,pecora1990synchronization,pikovsky2001universal}, yet real systems rarely achieve full coherence. Instead, cluster synchronization where subgroups of nodes synchronize independently-often prevails. { Cluster Synchronization} is linked to network symmetries and eigenvector centrality (EVC)~\cite{tudisco2021node}, with nodes in the same cluster sharing identical EVC values~\cite{zhang2020symmetry,khanra2022identifying,khanra2023endowing,mondal2024symmetry}. This connection between symmetry, EVC, and synchronization has deepened with studies on multilayer networks~\cite{della2020symmetries,sorrentino2020group}, where intra- and inter-layer interactions create intricate cluster patterns. Such systems better model real-world complexity, from social-technological interdependencies~\cite{de2013mathematical} to neuronal-immune cross-talk~\cite{chakravartula2017emergence}, necessitating frameworks that bridge dynamics and structure.

A paradigm shift emerged with the discovery that heterogeneity, rather than uniformity, can enhance synchronization. Motter et al.~\cite{nishikawa2003heterogeneity,motter2005network} demonstrated that parameter mismatches broaden synchronization regimes, a principle extended to ``explosive synchronization" in degree-frequency correlated networks~\cite{gomez2011explosive,khanra2018explosive}. Kundu et al.~\cite{kundu2017transition,kundu2018perfect} showed that aligning natural frequencies with node degrees induces perfect synchronization, bypassing the limitations of uniform parameters. However, these advances largely focus on single-layer networks, leaving a critical gap: How does heterogeneity shape synchronization in multilayer systems with phase lag?

Phase-lag, inherent in systems like power grids with transmission delays~\cite{dorfler2011critical} or neuronal networks with synaptic plasticity~\cite{banerjee2012enhancing}, disrupts synchronization by introducing temporal misalignment. The Sakaguchi-Kuramoto (SK) model~\cite{kuramoto1984chemical,acebron2005kuramoto} incorporates this lag ($\alpha$), often suppressing { global synchronization} except in specialized frequency configurations~\cite{kundu2018perfect}. Yet, in multilayer networks-where lag and interlayer coupling interact- { cluster synchronization} can persist in specific parameter regimes~\cite{nicosia2013remote,vlasov2017hub}. The stability of such states remains poorly understood, particularly under heterogeneous frequency distributions.

In this paper, we bridge this gap by investigating how degree-correlated natural frequencies stabilize { cluster synchronization} in multilayer SK networks with phase-lag. We analyze four frequency distributions: (1) uni-uni (uniform frequencies in both layers), (2) deg-uni (degree-correlated in one layer, uniform in the other), (3) uni-deg (the inverse), and (4) deg-deg (degree-correlated in both). Using numerical simulations and analytical stability criteria, we demonstrate that deg-deg distributions outperform others, sustaining { cluster synchronization} even at high $\alpha$. Key contributions include some mechanistic insight where degree-frequency coupling counteracts phase-lag induced desynchronization by aligning structural and dynamical heterogeneities, stability framework where derivation of a master stability function has been done for multilayer networks and validated via transversal Lyapunov exponents (TLEs)~\cite{pecora1998master,dabrowski2012largest,liao2016novel}, and finally, providing some practical relevance for designing resilient systems, from delay-tolerant power grids~\cite{xu2015explosive} to adaptive neural networks~\cite{banerjee2017enhancing}.
Our results reveal that deg-deg networks exhibit extended synchronization plateaus whereas uni-uni systems fail. This resilience stems from the interplay between degree heterogeneity and lag compensation, a phenomenon absent in partially correlated or uniform cases. By unifying symmetry-based cluster analysis~\cite{pecora2014cluster} with stability theory, this work advances the design of robust multilayer systems operating under real-world constraints. 
{\section{Cluster of a Multilayer Network}\label{cluster}
In this article, we investigate the phenomenon of cluster synchronization in the Sakaguchi-Kuramoto model on multilayer networks. Before presenting the dynamical model in the next section, we first introduce the formal structure of a multilayer network and define the notion of a cluster within such a network.
Consider a multilayer network composed of $M$ layers, where the $\sigma$-th ($\sigma \in \{1,2,\dots,M\}$) layer contains $N_\sigma$ nodes. The {\it intralayer} adjacency matrix $A^{\sigma} \in \mathbb{R}^{N_{\sigma}\times N_{\sigma}}$  encodes the connectivity among the nodes within layer $\sigma$. For any two distinct layers $\sigma$ and $\sigma'$ $(\sigma, \sigma' \in \{1,2,\dots,M\})$, the {\it interlayer} adjacency matrix $A^{\sigma \sigma'} \in \mathbb{R}^{N_{\sigma}\times N_{\sigma'}}$ captures the connectivity between their nodes.

Each layer $\sigma~(\sigma =1,2, \dots, M)$ is described as a connected graph $G^{\sigma}({V^{\sigma}},{E}^{\sigma})$, where ${V^{\sigma}}$ and ${E}^{\sigma}$ denote the set of vertices and edges respectively, with $|{V}^{\sigma}|=N_{\sigma}$. The edge set ${E}^{\sigma}$ consists of all ordered pairs $(i,j) \in {V}^{\sigma}\times {V}^{\sigma}$ such that the $i^{th}$ and $j^{th}$ nodes (vertices) are connected i.e. ${E}^{\sigma}\subseteq {V}^{\sigma} \times {V}^{\sigma}$. On the other hand, each pair of layers $\sigma$ and $\sigma{'}$ ($\sigma\neq \sigma^{'}$) of the multilayer network is described by an interlayer bipartite graph $G^{\sigma \sigma^{'}}({V}^{\sigma}, {V}^{\sigma^{'}}, {E}^{\sigma \sigma^{'}})$, where ${E}^{\sigma \sigma^{'}} \subseteq {V}^{\sigma} \times {V}^{\sigma^{'}}$. The supra-adjacency matrix $A$ of dimension $N = \sum_{\sigma} N_{\sigma}$ for the multilayer network is then given by
\begin{eqnarray}
A=
\begin{bmatrix}
A^{1} & A^{12} & \cdots\\
A^{21} & A^{2} & \cdots\\
\vdots & \vdots & \ddots
\end{bmatrix}.
\end{eqnarray}
We assume that all edges are undirected, hence each intralayer matrix is symmetric: $A^{\sigma} = (A^{\sigma})^T$, and each interlayer matrix satisfies $A^{\sigma \sigma^{'}} = (A^{\sigma \sigma^{'}})^T$ for  $\sigma,\sigma^{'}=1,2,\dots,M$ ($\sigma \neq \sigma^{'}$), ensuring that the overall supra-adjacency matrix is symmetric: $A=A^{T}$. The multilayer network can then be described by the graph $\mathbf{G}(\mathbf{V}, \mathbf{E})$, where $\mathbf{V} = \bigcup\limits_{\sigma=1}^{M}V^{\sigma}$ and $\mathbf{E} = \bigcup\limits_{\sigma,\sigma^{'}=1}^{M}E^{\sigma\sigma^{'}}$, where $E^{\sigma\sigma}=E^{\sigma}.$}

The graph $G^{\sigma}$ is said to possess symmetry if there exists a bijective mapping $\Pi^{\sigma}: {V}^{\sigma} \rightarrow {V}^{\sigma}$ that preserves the adjacency relation of $G^{\sigma}$ in layer $\sigma$, i.e. the bijection $\Pi^{\sigma}$ relation is an automorphism for $G^{\sigma}$. In matrix terms, this corresponds to the existence of a permutation matrix $P^{\sigma}=P^{\sigma}(\Pi^{\sigma})$ such that $P^{\sigma} A^{\sigma} (P^{\sigma})^{-1} = A^{\sigma}$. Collection of all such $P^{\sigma}$ matrices forms the symmetry group of the graph $G^{\sigma}$ under matrix multiplication.  At the multilayer level, any matrix can be represented by a block-diagonal matrix of the form:

\begin{eqnarray}
P=
\begin{bmatrix}
P^1 & 0 & \cdots &0\\
0 & P^2 & \cdots &0\\
\vdots & \vdots & \ddots&\vdots\\
0&0&\dots & P^M
\end{bmatrix},\label{Permu_mat_A}
\end{eqnarray}\\
where each $P^\sigma$ is an element of the symmetry groups of the graph $G^\sigma$, and the block structure must satisfy the interlayer symmetry conditions:
\begin{equation}
P^{\sigma} A^{\sigma\sigma^{'}} = A^{\sigma\sigma^{'}} P^{\sigma^{'}} ~~~~\rm{and}~~~~ P^{\sigma^{'}} A^{\sigma^{'}\sigma} = A^{\sigma^{'}\sigma} P^{\sigma},~~~~\rm{for~all} ~\sigma \neq \sigma^{'}.\label{Conjugacy_rel}
\end{equation}

The collection of all such block-permutation matrices $P$ forms the symmetry group of the full multilayer graph $\mathbf{G}$ with respect to matrix multiplication. Now, the set of nodes $\mathbf{V}$ of the multilayer network is partitioned into disjoint invariant subsets  by looking at the action of all symmetry elements on $\mathbf{V}$.  Each such subset of nodes forms a `cluster' of the multilayer network. Remarkably, within each cluster, all nodes share the same eigenvector centrality of the supra-adjacency matrix $A$ ~\cite{khanra2022identifying}.

\begin{figure}
\includegraphics[height=!,width=0.75\textwidth]{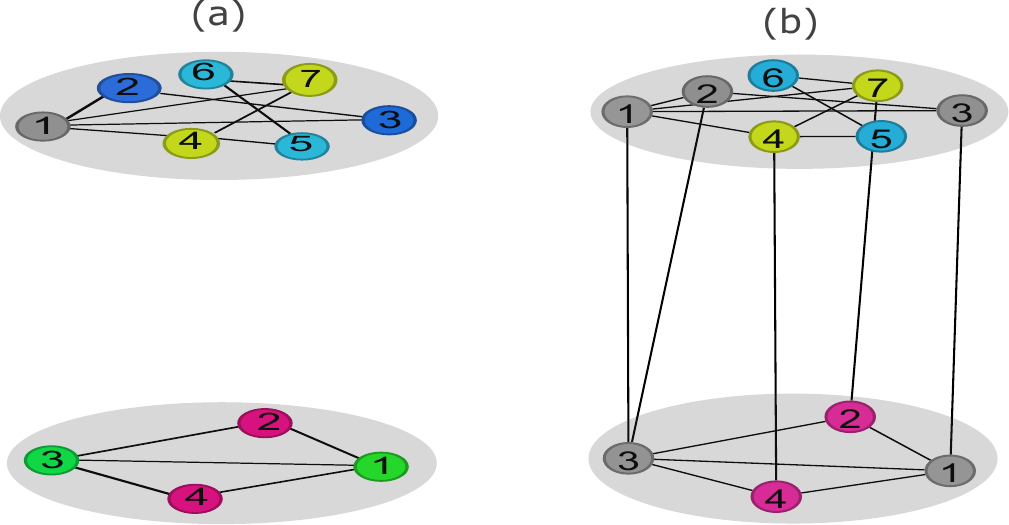}
\caption{{\bf A simple multilayer network diagram consisting of two layers.} In the absence of inter-layer connections(FIG(a)),
symmetry analysis confirms the clusters \{(2 3), (4 7), (5 6)\} in the layer 1 and \{(1 3), (2 4)\} in the layer 2. In the presence of
interlinks((FIG(b)), the orbit \{(2 3)\} of the layer 1 and \{(1 3)\} of the layer 2 are broken.} 
\label{Fig1}
\end{figure}

To illustrate these concepts, we consider a simple two-layer network shown in FIG.\ref{Fig1}(b), where Layer 1 contains seven nodes and Layer 2 contains four as shown in FIG.\ref{Fig1}(a). The distinct clusters identified in the absence and presence of interlayer links are highlighted using matching colors in FIG.\ref{Fig1}(a) and FIG.\ref{Fig1}(b) respectively.
It is important to note that the set of clusters in the multilayer network is not simply the union of the cluster sets of the individual layers when interlayer links are absent. The Adjacency matrices $A^1$ and $A^2$ respectively associated with the layers 1 and 2 in the absence of interlayer connections, along with their eigenvector centrality (EVC), are presented below.
\begin{eqnarray}
	A^1=
	\begin{pmatrix}
	\colorbox{red!40}0 & \colorbox{red!40}1 & \colorbox{red!40}1 & \colorbox{red!40}1 & \colorbox{red!40}0 & \colorbox{red!40}0 & \colorbox{red!40}1\\
	\colorbox{red!40}1 & \colorbox{red!40}0 & \colorbox{red!40}1 & \colorbox{red!40}0 & \colorbox{red!40}0 & \colorbox{red!40}0 & \colorbox{red!40}0\\
	\colorbox{red!40}1 & \colorbox{red!40}1 & \colorbox{red!40}0 & \colorbox{red!40}0 & \colorbox{red!40}0 & \colorbox{red!40}0 & \colorbox{red!40}0\\
	\colorbox{red!40}1 & \colorbox{red!40}0 & \colorbox{red!40}0 & \colorbox{red!40}0 & \colorbox{red!40}1 & \colorbox{red!40}0 & \colorbox{red!40}1\\
	\colorbox{red!40}0 & \colorbox{red!40}0 & \colorbox{red!40}0 & \colorbox{red!40}1 & \colorbox{red!40}0 & \colorbox{red!40}1 & \colorbox{red!40}0\\
	\colorbox{red!40}0 & \colorbox{red!40}0 & \colorbox{red!40}0 & \colorbox{red!40}0 & \colorbox{red!40}1 & \colorbox{red!40}0 & \colorbox{red!40}1\\
	\colorbox{red!40}1 & \colorbox{red!40}0 & \colorbox{red!40}0 & \colorbox{red!40}1 & \colorbox{red!40}0 & \colorbox{red!40}1 & \colorbox{red!40}0\\
	\end{pmatrix},~~
	\textbf{EVC}=
	\begin{pmatrix}
	\color{gray}0.540268167514888\\
	\color{blue}0.304254241202129\\
	\color{blue}0.304254241202129\\
	\color{yellow}0.445560407855931\\
	\color{cyan}0.250919176721960\\
	\color{cyan}0.250919176721959\\
	\color{yellow}0.445560407855931
	\end{pmatrix}
	\label{Fig1upperlayer}
\end{eqnarray}

\begin{eqnarray}
	A^2=
	\begin{pmatrix}
	\colorbox{blue!40}0 & \colorbox{blue!40}1 & \colorbox{blue!40}1 & \colorbox{blue!40}1\\
	\colorbox{blue!40}1 & \colorbox{blue!40}0 & \colorbox{blue!40}1 & \colorbox{blue!40}0\\
	\colorbox{blue!40}1 & \colorbox{blue!40}1 & \colorbox{blue!40}0 & \colorbox{blue!40}1\\
	\colorbox{blue!40}1 & \colorbox{blue!40}0 & \colorbox{blue!40}1 & \colorbox{blue!40}0
	\end{pmatrix},~~
	\textbf{EVC}=
	\begin{pmatrix}
	\color{green}0.557345410189304\\
	\color{magenta}0.435162146493599\\
	\color{green}0.557345410189304\\
	\color{magenta}0.435162146493599
	\end{pmatrix}
	\label{Fig1lowelayer}
\end{eqnarray}
The supra adjacency matrix  in the presence of interlayer links and corresponding EVC are then obtained as
\begin{eqnarray}
	A=
	\begin{pmatrix}
	\colorbox{red!40}0 & \colorbox{red!40}1 & \colorbox{red!40}1 & \colorbox{red!40}1 & \colorbox{red!40}0 & \colorbox{red!40}0 & \colorbox{red!40}1 & 0 & 0 & 1 & 0\\
	\colorbox{red!40}1 & \colorbox{red!40}0 & \colorbox{red!40}1 & \colorbox{red!40}0 & \colorbox{red!40}0 & \colorbox{red!40}0 & \colorbox{red!40}0 & 0 & 0 & 1 & 0\\
	\colorbox{red!40}1 & \colorbox{red!40}1 & \colorbox{red!40}0 & \colorbox{red!40}0 & \colorbox{red!40}0 & \colorbox{red!40}0 & \colorbox{red!40}0 & 1 & 0 & 0 & 0\\
	\colorbox{red!40}1 & \colorbox{red!40}0 & \colorbox{red!40}0 & \colorbox{red!40}0 & \colorbox{red!40}1 & \colorbox{red!40}0 & \colorbox{red!40}1 & 0 & 0 & 0 & 1\\
	\colorbox{red!40}0 & \colorbox{red!40}0 & \colorbox{red!40}0 & \colorbox{red!40}1 & \colorbox{red!40}0 & \colorbox{red!40}1 & \colorbox{red!40}0 & 0 & 0 & 0 & 0\\
	\colorbox{red!40}0 & \colorbox{red!40}0 & \colorbox{red!40}0 & \colorbox{red!40}0 & \colorbox{red!40}1 & \colorbox{red!40}0 & \colorbox{red!40}1 & 0 & 0 & 0 & 0\\
	\colorbox{red!40}1 & \colorbox{red!40}0 & \colorbox{red!40}0 & \colorbox{red!40}1 & \colorbox{red!40}0 & \colorbox{red!40}1 & \colorbox{red!40}0 & 0 & 1 & 0 & 0\\
	0 & 0 & 1 & 0 & 0 & 0 & 0 & \colorbox{blue!40}0 & \colorbox{blue!40}1 & \colorbox{blue!40}1 & \colorbox{blue!40}1\\
	0 & 0 & 0 & 0 & 0 & 0 & 1 & \colorbox{blue!40}1 & \colorbox{blue!40}0 & \colorbox{blue!40}1 & \colorbox{blue!40}0\\
	1 & 1 & 0 & 0 & 0 & 0 & 0 & \colorbox{blue!40}1 & \colorbox{blue!40}1 & \colorbox{blue!40}0 & \colorbox{blue!40}1\\
	0 & 0 & 0 & 1 & 0 & 0 & 0 & \colorbox{blue!40}1 & \colorbox{blue!40}0 & \colorbox{blue!40}1 & \colorbox{blue!40}0
	\end{pmatrix} \label{multi_Fig1},~~
	\textbf{EVC} = 
	\begin{pmatrix}
	\color{gray}0.423507576861891\\
	\color{gray}0.301072065845225\\
	\color{gray}0.281624982046096\\
	\color{yellow}{0.292166082614229}\\
	\color{cyan}0.105377314439929\\
	\color{cyan}0.105377314439929\\
	\color{yellow}{0.292166082614229}\\
	\color{gray}0.337870588189062\\
	\color{magenta}0.281166306945458\\
	\color{gray}0.430683175724854\\
	\color{magenta}0.281166306945458
	\end{pmatrix}.\label{Supra_adja}		
\end{eqnarray}
From the color-coded EVCs of the individual layers without interlayer links, as well as those of the complete multilayer network, we observe that the coloring of the EVC elements aligns well with the cluster structures identified through the preceding symmetry analysis in each case. We now proceed to introduce the generalized Sakaguchi-Kuramoto (SK) model on top of the multilayer network, which is discussed in detail in the following section.

\section{Model Description}
For investigating cluster synchronization on multilayer networks, we consider a generalized version of the { Sakaguchi-Kuramoto (SK)} model on top of the network. In the network, the phase $\theta_i^{\sigma}$ of node $i (=1,\dots,N^{\sigma})$ in layer $\sigma (=1,\dots,M)$ of the multilayer network is governed by the coupled differential equation

\begin{eqnarray}\label{Model_eqn}
\dot \theta_i^{\sigma} = \omega_i^{\sigma} + \epsilon^{\sigma}\sum_{j=1}^{\mathrm{N^{\sigma}}}A_{ij}^{\sigma}\sin(\theta_j^{\sigma} - \theta_i^{\sigma}-\alpha^{\sigma})  
+ \sum_{\sigma \neq \sigma^{'}}\epsilon^{\sigma \sigma^{'}} \sum_{j=1}^{\mathrm{N^{\sigma^{'}}}}A_{ij}^{\sigma \sigma^{'}}\sin(\theta_j^{\sigma^{'}}-\theta_i^{\sigma}),
\end{eqnarray}
where $A_{ij}^{\sigma}$ and $\omega_i^{\sigma}$ respectively denote the $ ij^{th}$ element of the adjacency matrix and natural frequency of the $i$th oscillator in layer $\sigma$, while $\epsilon^{\sigma}$ and $\alpha^{\sigma}$ denote the uniform coupling strength and phase-lag parameters respectively for layer $\sigma$. $\epsilon^{\sigma \sigma^{'}}$ and $A_{ij}^{\sigma \sigma^{'}}$ indicate interlayer coupling strength and the $ij^{th}$ element of the interlayer coupling matrix respectively between the layers $\sigma$ and $\sigma^{'}$ $(\sigma \neq \sigma ^{'})$.

To measure the cluster synchronization in the network, we introduce the cluster order parameter $R_{\mathrm{cluster}}^{\sigma}$ for the entire population of layer $\sigma$ defined by
\begin{eqnarray}\label{R_cluster}
 R_{\mathrm{cluster}}^{\sigma} =
\frac{1}{\wp ^{\sigma}}\sum_{k=1}^{\wp ^{\sigma}}R_k^{\sigma}, 
\end{eqnarray}
where  $\wp^{\sigma}$ is the total number of clusters (including trivial and non-trivial clusters) and $R_k^{\sigma}$ is the global order parameter of the $k$th cluster in layer $\sigma$ which is again defined as
\begin{eqnarray}\label{Global_R}
 R_k^{\sigma} e^{i(\psi^\sigma-\alpha^\sigma)} =
\frac{1}{N_k^{\sigma}}\sum_{j\in v_k^{\sigma}}e^{i(\theta_j^{\sigma}-\alpha^{\sigma})}, 
\end{eqnarray}
where $N_k$ is the number of nodes in the $k$th cluster and $v_k$ is the set of nodes involved in the cluster $k$. ${R_k}^{\sigma}=1$ quantifies a perfectly synchronized $k$th cluster and ${0}\leq{R_k}^{\sigma}<{1}$ implies desynchronized cluster.   
This paper aims to elucidate the intricate relationship between frequency heterogeneity, phase-lag, and cluster synchronization in multilayer networks, providing valuable insights into the dynamics governing complex systems.

\begin{figure*}
\centering
\includegraphics[ width=0.8\linewidth]{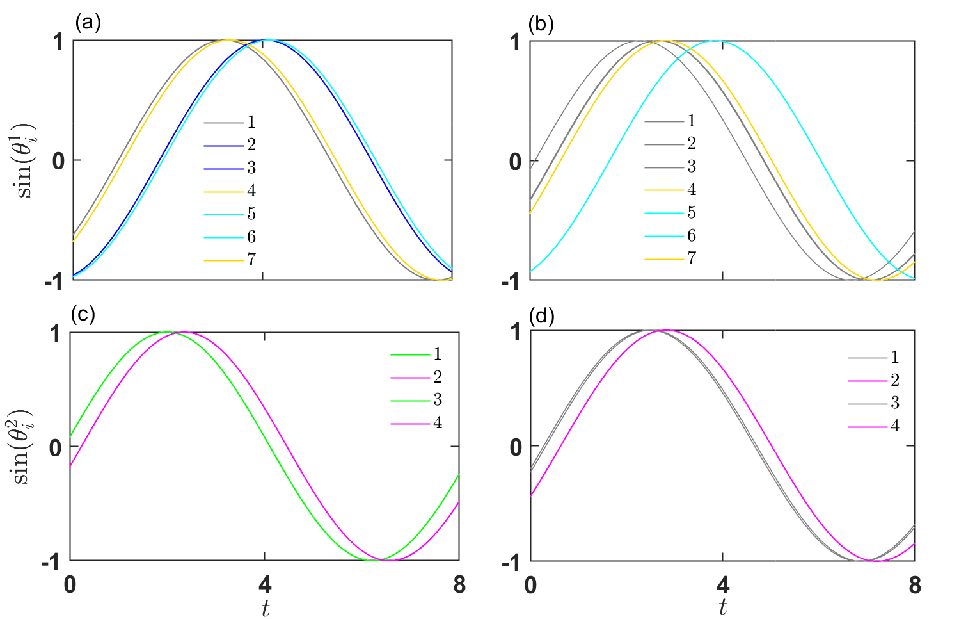}
\caption{{\bf Time series of the network described in FIG.\ref{Fig1} in the presence and absence of interlayer interactions}. First and second rows correspond to the layers 1 and 2 respectively. The first column ((a) and (c)) shows the time series in absence of inter-layer coupling, while that in the presence of the inter-layer coupling are shown in the second column ((b) and (d)). For numerical simulation we take the parameter as $\alpha^{\sigma} = 0.8, ~\epsilon^{\sigma} = 1 $ and $\epsilon^{\sigma\sigma'} = 1$. { Here, the curve number corresponds to the color matched nodes in FIG.\ref{Fig1}.} From this figure, it is seen that the { time series for the nodes in synchronized clusters merged together, which is} consistent with the symmetry as well as EVC analysis.}
\label{Figure2}
\end{figure*}

\section{Cluster synchronization of Sakaguchi-Kuramoto model in Multilayer Network }

We now consider { Sakaguchi-Kuramoto} dynamics over the top of the multilayer network illustrated in the FIG.~\ref{Fig1}
with uniform natural frequency for all nodes in both layer ($\omega_i =1$, $\omega_i =1$). Under this setting, the system exhibits cluster synchronization across a wide range of the phase-lag parameter $\alpha$ ($0\leq \alpha \leq 1.12$). Numerical simulations reveal that the coupled { Sakaguchi-Kuramoto} oscillators on the multilayer network evolve into a statistically stationary state where synchronized clusters are formed by the set of nodes belonging to a cluster of the multilayer network as defined in section~\ref{cluster}.  These clusters correspond to groups of nodes as defined in section~\ref{cluster}, and importantly, match the symmetry-based cluster structures determined from the eigenvector centrality (EVC) framework proposed in~\cite{khanra2023endowing}. Within each cluster, all oscillators exhibit identical instantaneous phases.

The results of the numerical simulations performed with $\epsilon^{\sigma}=1,~(\sigma=1,2)$ and $\epsilon^{12}=1$ are shown in the FIG.\ref{Figure2}. FIG.\ref{Figure2}((a) and (b)) display the time evaluation of oscillator phases in each layer in the absence of interlayer coupling, while FIG.\ref{Figure2}((c) and (d)) show the corresponding results in presence of interlayer coupling. In both cases, the observed cluster synchronization is consistent with the clusters determined using the proposed theory. We find that, in the absence of inter layer coupling, the system reaches a stationary state after a sufficient transient period. The phase trajectories in the first layer organize into four distinct phase groups: $\{\theta_1^1\}$, $\{\theta_2^1, \theta_3^1\}$, $\{\theta_4^1, \theta_7^1\}$, and $\{\theta_5^1, \theta_6^1\}$, satisfying
\begin{equation}
\theta_2^1(t)=\theta_3(t)^1,~\theta_4(t)^1=\theta_7(t)^1,~\mathrm{and}~\theta_5(t)^1=\theta_6(t)^1. \label{eqn10}
\end{equation} 
Meanwhile, the second layer forms two phase groups: $\{\theta_1^2, \theta_3^2\}$ and $\{\theta_2^2, \theta_4^2\}$, satisfying 
\begin{equation}
~\theta_1(t)^2=\theta_3(t)^2,~\mathrm{and}~\theta_2(t)^2=\theta_4(t)^2.\label{eqn11}
\end{equation}
In presence of interlayer coupling, the system exhibits three nontrivial clusters along with five trivial clusters shown in FIG.\ref{Figure2}((c) and (d)), satisfying 
\begin{eqnarray}
\theta_4(t)^1&=&\theta_7(t)^1,~\theta_5(t)^1=\theta_6(t)^1\nonumber\\
&&~\mathrm{and}~\theta_2(t)^2=\theta_4(t)^2.\label{eqn12}
\end{eqnarray}

These numerical simulation results further validate the theoretical framework by demonstrating consistent cluster synchronization behavior.

\begin{figure*}
\includegraphics[ width=1\linewidth]{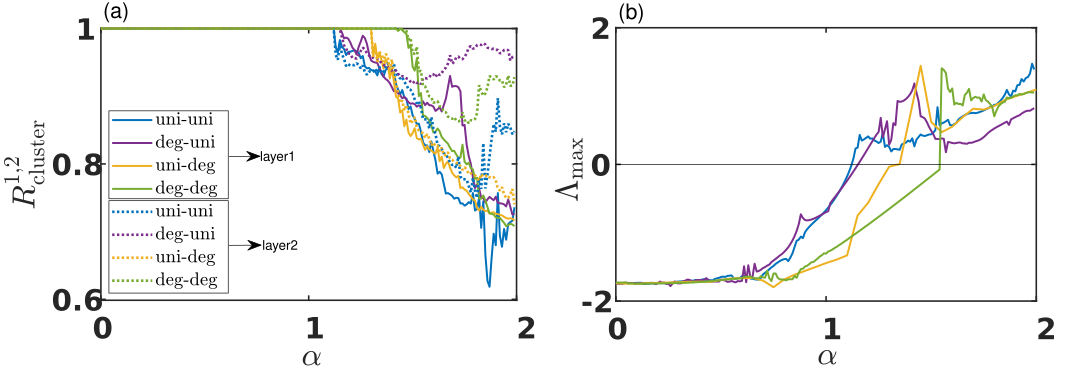}
\caption{Panel (a) shows the global order parameter computed from the cluster synchronization states in both layers as a function of the phase-lag parameter $\alpha$. Panel (b) presents the transversal Lyapunov exponent (TLE) for the network structure depicted in Fig.~\ref{Fig1}(b), providing a stability analysis of the identified cluster states. The coupling strength is fixed at $\epsilon=1$. The cluster order parameter $R_{\mathrm {cluster}}^{1,2}$ of each layer deviates from 1 around $\alpha_c \approx 1.12$, $1.15$, $1.3$, and $1.45$ which closely matched with the critical value of $\alpha$ in panel (b) where $\Lambda_{\rm max}$ changes it's sign.}
\label{Figure3}
\end{figure*}

To investigate the effect of frequency heterogeneity on cluster synchronization ($R_{\mathrm{cluster}}^{\sigma}$) with varying phase-lag, we explore different types of natural frequency distributions shown in FIG.\ref{Figure3}. In the first case, referred to as uni-uni configurations, both layers have uniform frequencies ($\omega_i=1$ for all $i$ in both layers). In the second case, referred to as deg-uni, the first layer has degree-correlated natural frequencies ($\omega_i = d_i$), while the second layer maintains uniform natural frequencies ($\omega_i = 1$). Conversely, in the uni-deg case, the roles are reversed: the first layer has uniformly distributed natural frequencies ($\omega_i = 1$), while the second layer exhibits degree correlated frequencies ($\omega_i = d_i$). Lastly, in the deg-deg configuration, both layers have degree-correlated frequencies ($\omega_i = d_i$, for all $i$ in both layers). These configurations allow us to systematically assess the influence of frequency heterogeneity and interlayer dynamics on the emergence and regime enhancement for the synchronized clusters.

The numerical integration of the system described by Eq.(\ref{Model_eqn}) is carried out using the the fourth order Runge-Kutta (RK4) method with time step 0.01 for the above mentioned pairs of frequency distributions. For each case, we compute the cluster order parameter $R_{\mathrm{cluster}}^{\sigma}$ for both layers using Eq.(\ref{R_cluster}), while varying the phase-lag parameters $\alpha^{\sigma}$ within the range $[0, 2]$ in increments of $\Delta \alpha^{\sigma} = 0.01$, for $\sigma = 1,2$. FIG.\ref{Figure3}(a) shows the variation of the order parameters as a function of $\alpha$. The navy blue, orange, purple, and green curve represent the cluster order parameters for uni-uni, deg-uni, uni-deg and deg-deg cases respectively. Here, we have used same color for both the layers, since both the layers are desynchronized at same $\alpha$ value. As previously discussed, cluster synchronization is considered intact when $R_{\mathrm{cluster}}^{\sigma}=1$ and breaks down when $R_{\mathrm{cluster}}^{\sigma} < 1$.

It is evident from Fig.~\ref{Figure3}(a) that in the deg-deg case, cluster synchronization persists over a wider range of $\alpha$ values compared to the other scenarios. Specifically, the breakdown of synchronization (i.e., the point where $R_{\mathrm{cluster}}^{\sigma} < 1$) occurs at $\alpha_c \approx 1.12$, $1.15$, $1.3$, and $1.45$ for the uni-uni, deg-uni, uni-deg, and deg-deg distributions, respectively. This indicates that degree-frequency correlation enhances the robustness of cluster synchronization, likely due to the embedding of structural network information into the system's dynamics.
To further assess the stability of the synchronized clusters, we compute the transversal Lyapunov exponents (TLEs) using the master stability function approach~\cite{inubushi2023characterizing,andrieu2016transverse,dabrowski2012largest}, whose derivation is detailed in Subsection~\ref{stablity_analysis}. Figure~\ref{Figure3}(b) shows the variation of the TLEs with $\alpha$ for the four frequency distributions, with the same color scheme as in Fig.\ref{Figure3}(a). The TLEs (see Eq.(\ref{Variational_Eqn})) transition from negative to positive values at the same critical $\alpha$ values where $R_{\mathrm{cluster}}^{\sigma}$ drops below 1. This transition confirms the loss of cluster synchronization stability and validates our theoretical framework.

Finally, we evaluate the general applicability of our cluster analysis theory by testing it on several different multilayer network configurations, as illustrated in Fig.~\ref{Figure4}.

\subsection{Role of Heterogeneity in Cluster synchronization}
\begin{figure*}
\includegraphics[width=1\linewidth]{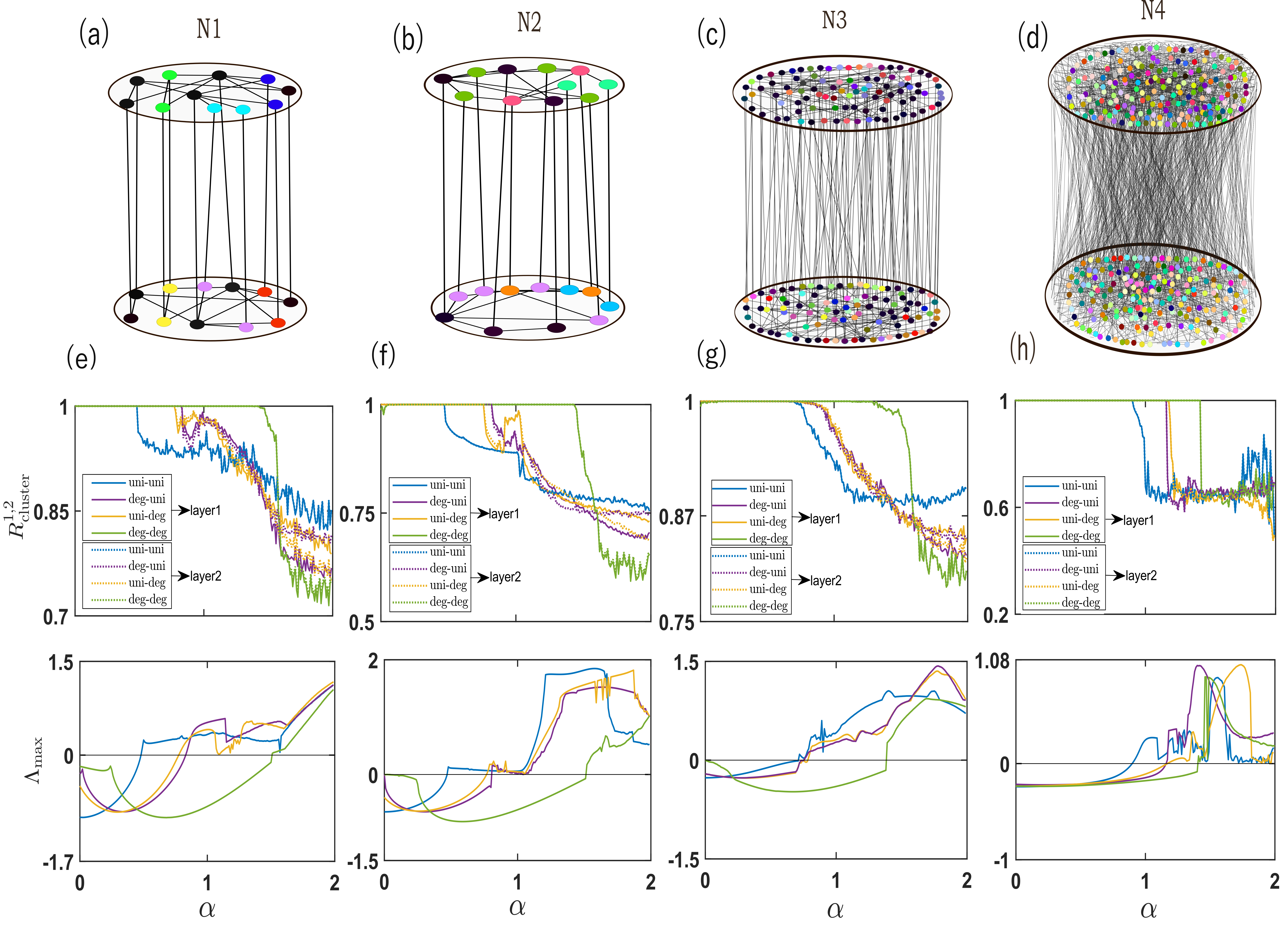}
\centering
\caption{(a–d) Four synthetic multilayer networks (N1–N4) visualized using Gephi and edited with Inkscape.
(a) Network N1 consists of 22 nodes organized into six nontrivial clusters: two blue, two light green, and two cyan clusters in the upper layer; and two magenta, two yellow, and two red clusters in the lower layer. The stability of these clusters is shown in panel (e) as a function of the phase-lag parameter $\alpha$, with fixed coupling strength $\epsilon = 1$. The cluster order parameter $R^{1,2}_{\rm cluster}$ for each layer deviates from 1 at critical values $\alpha_c \approx 0.48, 0.77, 0.82$, and $1.43$, closely matching the values at which the maximum Lyapunov exponent $\Lambda_{\rm max}$ changes sign for the following frequency combinations: uni-uni (navy blue), deg-uni (orange), uni-deg (purple), and deg-deg (green).
(b) Network N2 also has 22 nodes forming six nontrivial clusters: four green, two red, and two aqua in the upper layer; and four magenta, two orange, and two heavy cyan in the lower layer. The cluster stability is shown in panel (f) as a function of $\alpha$, with deviations of $R^{1,2}_{\rm cluster}$ from 1 around $\alpha_c \approx 0.47, 0.76, 0.82$, and $1.43$, again aligning with transitions in $\Lambda_{\rm max}$ for the same frequency combinations and color coding.
(c) Network N3 includes 200 nodes and 38 nontrivial clusters. The stability results are shown in panel (g), with critical values $\alpha_c \approx 0.70, 0.78, 0.78$, and $1.29$ for the four frequency configurations (uni-uni, deg-uni, uni-deg, and deg-deg).
(d) Network N4 comprises 400 nodes forming 200 nontrivial clusters. Panel (h) shows cluster stability with critical values $\alpha_c \approx 0.90, 1.16, 1.18$, and $1.34$ for the respective frequency combinations. All simulations were performed at a fixed coupling strength $\epsilon = 1$.}
\label{Figure4}
\end{figure*}

To investigate the impact of frequency heterogeneity on cluster synchronization (CS), we explored various natural frequency distributions while solving Eq.(\ref{Model_eqn}) and computed the cluster order parameter using Eq.(\ref{R_cluster}). Our findings reveal that CS is most robust when the frequencies are degree-degree (deg-deg) correlated, as illustrated in Fig.~\ref{Figure3}. We further examine the role of heterogeneity by applying the same frequency distributions to a collection of synthetically generated multilayer networks, illustrated in (see Fig.~\ref{Figure4}). Specifically, we considered four network topologies:
\begin{itemize}
    \item $\rm N_1$: A small network of 22 nodes, featuring 10 trivial clusters and 6 non-trivial clusters involving 12 nodes.
    \item $\rm N_2$: Also with 22 nodes, but organized into 6 trivial clusters and 6 non-trivial clusters encompassing 16 nodes.
    \item $\rm N_3$: A larger network with 200 nodes, containing 96 trivial clusters and 38 non-trivial clusters involving 104 nodes.
    \item $\rm N_4$: The largest network, composed of 400 nodes, with all clusters being non-trivial.
\end{itemize}
Across all these network configurations, the deg-deg frequency distribution consistently outperformed the other types, providing the most favorable conditions for sustaining and enhancing cluster synchronization. These findings underscore the importance of aligning structural properties with dynamical features to maintain coherent behavior in complex multilayer systems.

\begin{figure*}
\includegraphics[width=1.0\linewidth]{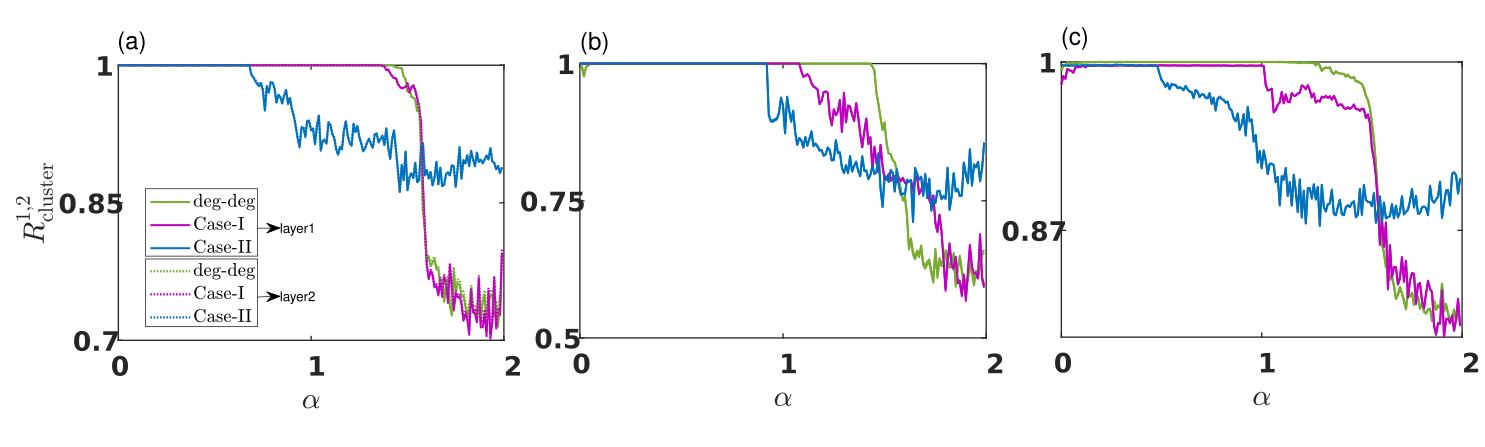}
\centering
\caption{\textbf{Figure Panel (a–c):} Cluster order parameter $R^{1,2}_{\mathrm{cluster}}$ as a function of the phase-lag parameter $\alpha$, computed for the networks N1, N2, and N3, respectively, as shown in FIG.~\ref{Figure5}(a–c). All three simulations were done by the fixed coupling strength at 1.} 
\label{Figure5}
\end{figure*}

Additionally, we analyzed two more nuanced frequency distribution schemes (see Fig.\ref{Figure5}): (i) in the first configuration, non-trivial cluster nodes were assigned uniform frequencies while trivial nodes followed a degree-correlated distribution (Case-I), and (ii) in the second configuration, non-trivial nodes had degree-correlated frequencies and trivial nodes had uniform frequencies (Case-II). Even in these hybrid cases, the deg-deg correlated distribution outperformed both the Case-I and Case-II configurations except in the case of small-sized networks (e.g., the network shown in Fig.\ref{Fig1}(b)), where performance differences across frequency configurations were less pronounced due to limited structural complexity. It is worth noting that the largest network in our study, $\rm N_4$, was excluded from this hybrid frequency analysis, as it contains only non-trivial clusters and no trivial ones, making these hybrid schemes inapplicable.

\subsection{Stability Analysis of cluster synchronized state}\label{stablity_analysis}
The model Eq.(\ref{Model_eqn}) of the main text can be written as,\\
\begin{eqnarray}\label{General_Model_eqn}
\dot \theta_i^{\sigma} = f(\theta_i^{\sigma})+\epsilon^{\sigma}\sum_{j=1}^{\mathrm{N^{\sigma}}}A_{ij}^{\sigma} \mathcal{H}^{\sigma \sigma}(\theta_j^{\sigma}, \theta_i^{\sigma})
+\sum_{\sigma \neq \sigma^{'}}\epsilon^{\sigma \sigma^{'}} \sum_{j=1}^{\mathrm{N^{\sigma^{'}}}}A_{ij}^{\sigma \sigma^{'}}\mathcal{H}^{\sigma \sigma'}(\theta_j^{\sigma'}, \theta_i^{\sigma}),
\end{eqnarray}

where $\mathcal{H}^{\sigma \sigma}$ denotes the underlying coupling function between the nodes of layer $\sigma$ and $\mathcal{H}^{\sigma \sigma'}$ denotes the coupling function between the layers $\sigma, \sigma'$. In our case we have taken sinusoidal coupling function with phase frustration in intralayer coupling.

Using the EVC framework, layer $\sigma$ can be partitioned into $\wp^{\sigma}$ number of orbital clusters namely $\mathcal{C}^\sigma_1,~\mathcal{C}^\sigma_2,\dots,\mathcal{C}^\sigma_{\wp^\sigma}$.\\
Let us define a $\wp^{\sigma}\times\wp^{\sigma}$ intralayer quotient matrix $\mathcal{Q}^{\sigma \sigma}$ such that for each pair of $\sigma$~-~ clusters $(\mathcal{C}^\sigma_u,\mathcal{C}^\sigma_v)$,\\
\begin{eqnarray}
\mathcal{Q}^{\sigma \sigma}_{uv}= \sum_{j\in \mathcal{C}^\sigma_v}A_{ij}^{\sigma}~;~~~~~\forall i\in\mathcal{C}^\sigma_u,~~~u,v=1,2,\dots,\wp^{\sigma},
\end{eqnarray}

and the corresponding $\wp^{\sigma}\times\wp^{\sigma'}$ interlayer quotient matrix $\mathcal{Q}^{\sigma \sigma^{'}}$ can be written in the form such that for each pair of clusters, $\mathcal{C}^\sigma_u$ will be from the layer $\sigma$ and the second cluster $\mathcal{C}^{\sigma^{'}}_v$ will be selected from layer $\sigma^{'}$~ $(\sigma \neq \sigma^{'})$ and is given by,\\
\begin{eqnarray}
\mathcal{Q}^{\sigma \sigma^{'}}_{uv}= \sum_{j \in \mathcal{C}^{\sigma^{'}}_v}A_{ij}^{\sigma \sigma^{'}}~;~~~\forall i\in\mathcal{C}^\sigma_u,~~~u=1,2,\dots,\wp^{\sigma}~~\&~~~~,\\~v=1,2,\dots,\wp^{\sigma'}\nonumber.
\end{eqnarray}
Let $\mathcal{S}^{\sigma}_u$ indicates the synchronized manifold for the clusters $\mathcal{C}^\sigma_u$ in layer $\sigma$,~$u=1,2,\dots,\wp^{\sigma}$. Thus the time evolution equation for the synchronized manifold can be written as,\\
\begin{eqnarray}\label{Synchronized_manifold}
\dot {\mathcal{S}}^{\sigma}_u = f({\mathcal{S}}^{\sigma}_u)+\epsilon^{\sigma}\sum_{v=1}^{\wp^{\sigma}}\mathcal{Q}_{uv}^{\sigma \sigma} \mathcal{H}^{\sigma \sigma}(\mathcal{S}^{\sigma}_v, \mathcal{S}^{\sigma}_u)
+\sum_{\sigma \neq \sigma^{'}}\epsilon^{\sigma \sigma^{'}} \sum_{v=1}^{\wp^{\sigma^{'}}}\mathcal{Q}_{uv}^{\sigma \sigma^{'}}\mathcal{H}^{\sigma \sigma'}(\mathcal{S}^{\sigma^{'}}_v, \mathcal{S}^{\sigma}_u) ,
\end{eqnarray}
where $\sigma=1,2,\dots,M$ and $u=1,2,\dots,\wp^{\sigma}$. Every node $\theta_i^{\sigma}$ of layer $\sigma$ is mapped into one synchronized manifold $\mathcal{S}^{\sigma}_u$,~$u=1,2,\dots,\wp^{\sigma}$ such that $\theta_i^{\sigma}\equiv \mathcal{S}^{\sigma}_u$, $\forall i \in \mathcal{C}^{\sigma}_u$. Now considering small perturbations $\delta{\theta_i^{\sigma}}$ around the synchronized manifold, the phase becomes $\theta_i^{\sigma}=\mathcal{S}^{\sigma}_u+\delta{\theta_i^{\sigma}}$, $\forall i \in \mathcal{C}^{\sigma}_u$. Then the variational equation corresponding to Eq.(\ref{General_Model_eqn}) is,\\
\begin{eqnarray}\label{Variational_Eqn}
\dot \delta \theta^{\sigma}_i = \mathbf{J} f({\mathcal{S}}^{\sigma}_u) \delta \theta^{\sigma}_i+ \epsilon^{\sigma}\sum_{j=1}^{N^{\sigma}} A_{ij}^{\sigma \sigma} \mathbf{J}\mathcal{H}^{\sigma \sigma}(\mathcal{S}^{\sigma}_v, \mathcal{S}^{\sigma}_u)\delta \theta^{\sigma}_j \delta \theta^{\sigma}_i
+\sum_{\sigma \neq \sigma^{'}}\epsilon^{\sigma \sigma^{'}} \sum_{j=1}^{N^{\sigma^{'}}} A_{ij}^{\sigma \sigma^{'}} \mathbf{J}\mathcal{H}^{\sigma \sigma^{'}}(\mathcal{S}^{\sigma^{'}}_v, \mathcal{S}^{\sigma}_u)\delta \theta^{\sigma^{'}}_j \delta \theta^{\sigma}_i,
\end{eqnarray}
where $\mathbf{J}f$, $\mathbf{J}\mathcal{H}^{\sigma \sigma}$ and $\mathbf{J}\mathcal{H}^{\sigma \sigma^{'}}$ denote the jacobian corresponding to $f$, the intralayer coupling function $\mathcal{H}^{\sigma \sigma}$ and the interlayer coupling function $\mathcal{H}^{\sigma \sigma^{'}}$. Equation(\ref{Variational_Eqn}) is the required master stability function to check the stability of the cluster states. Now we calculate the maximal Lyapunov exponent ($\Lambda_{\rm max}$) of the clusters along the transverse direction using the Eq.(\ref{Synchronized_manifold}).The negative value of that transversal Lyapunov exponent indicates the stable region for the clusters, and the positive value indicates the unstable region. 
We have calculated  the stability of cluster states of coupled phase-frustrated oscillators for five different networks shown in FIG.\ref{Figure4} (e-h). The networks and associated clusters are shown in  FIG.\ref{Figure4} (a-d). Note that all the cluster nodes are identified with eigenvector centrality of the supra-adjacency matrix $\mathcal{A}$. Using the stability analysis of coupled phase oscillators we have identified a wide range of $\alpha$ where the synchronous cluster states are stable FIG.\ref{Figure4} (e-h).

\section{CONCLUSIONS}

In this study, we have investigated the emergence and stability of cluster synchronization (CS) in multilayer networks of phase oscillators governed by the Sakaguchi-Kuramoto model, with particular emphasis on scenarios involving significant phase-lag. Our findings highlight the critical role played by the distribution of natural frequencies in determining the robustness and persistence of cluster synchronization. We systematically explored a variety of frequency distributions, including uni-uni, deg-uni, uni-deg, and deg-deg, and our results show that the deg-deg distribution significantly enhances cluster synchronization compared to the others. This configuration allows the structural heterogeneity of the network, encoded via node degrees, to be reflected in the oscillator dynamics, thereby promoting more coherent and stable clustering across layers.
We further extended our analysis to synthetically generated multilayer networks with both trivial and non-trivial cluster structures. Even in cases where only a subset of nodes formed non-trivial clusters, the deg-deg distribution consistently preserved synchronization over a wider range of phase-lag parameters. Additionally, we examined hybrid frequency distributions, where trivial and non-trivial nodes followed different rules for frequency assignment. Again, the deg-deg configuration showed superior performance, especially in larger networks, suggesting that structural-dynamical alignment is crucial for sustaining { cluster synchronization} under realistic and heterogeneous conditions.
Our findings contribute to a deeper understanding of how intrinsic frequency heterogeneity influences the collective dynamics of multilayer systems that experience phase lags. These insights are particularly relevant for the design and control of synchronization in real-world networks where nodes differ in both structural importance and dynamical behavior.

Looking ahead, several promising research directions emerge from this work. First, our approach can be extended to higher-order Kuramoto models, where multi-node interactions may offer richer dynamics and new forms of synchronization. Second, we plan to explore various centrality measures such as betweenness, closeness, or eigenvector centrality, as internal node frequencies to identify the optimal distribution for enhancing { cluster synchronization}. Third, the impact of partial control strategies, where only a fraction of nodes in one layer are assigned degree-based frequencies while others maintain uniform frequencies, offers another avenue for selectively enhancing synchronization while minimizing external intervention.

In summary, this study lays a solid foundation for leveraging frequency-structure correlations to foster robust synchronization in complex multilayer networks. These principles may guide future efforts in network design, control, and analysis across disciplines ranging from neuroscience and engineering to social dynamics and beyond.

\section{Declaration of generative AI and AI-assisted technologies in the writing process}
During the preparation of this work the authors used language refinement tools in order to refine the english style of the presentation alongside the grammatical corrections. After using this tool/service, the authors reviewed and edited the content as needed and takes full responsibility for the content of the publication.
   
\bibliographystyle{apsrev4-1}
\bibliography{Ref}
\end{document}